\documentclass[review]{elsarticle}
\usepackage{lineno,hyperref}
\modulolinenumbers[4]
\journal{Journal of \LaTeX\ Templates}

\usepackage{amsmath}
\usepackage{mathrsfs}
\usepackage{subfigure}
\usepackage{graphicx}

\biboptions{authoryear}

\begin{document}
\begin{frontmatter}
\title{A fast point-pattern matching algorithm based on statistical method }
\author[address1,address2,address3]{Zhen-Jun Zhang\corref{mycorrespondingauthor}}
\ead{zjzhang@ynao.ac.cn}
\cortext[mycorrespondingauthor]{Corresponding author}
\author[address1,address2,address3]{Yi-Gong Zhang}
\author[address1,address2]{Xiang-Ming Cheng}
\author[address1,address2]{Jian-Cheng Wang}

\address[address1]{Yunnan Observatories, Chinese Academy of Sciences, Kunming 650216, China}
\address[address2]{Key Laboratory of the Structure and Evolution of Celestial Objects, Chinese Academy of Sciences, Kunming 650216, China}
\address[address3]{University of Chinese Academy of Sciences, Beijing 100049, China}

\begin{abstract}
We propose a new pattern-matching algorithm for matching CCD images to a stellar catalogue based statistical method in this paper. The method of constructing star pairs can greatly reduce the computational complexity compared with triangle method.  We use a subsample of the brightest objects from the image and reference catalogue, and find a coordinate transformation between the image and reference catalogue based on the statistical information of star pairs. Then all the objects are matched based on the initial plate solution.
The matching process can be accomplished in several milliseconds for the observed images taken by Yunnan observatory 1-m telescope .
\end{abstract}

\begin{keyword}
\texttt{astrometry, pattern-matching algorithm, Statistical approach}
\end{keyword}

\end{frontmatter}

\linenumbers

\section{Introduction}
A fast and robust pattern-matching method is very important for the modern astronomical observation and spacecraft attitude control based on star sensor. Especially for  the course of the wide-field CCD imaging survey or high speed image acquisition detector such as EMCCD and SCMOS camera applied in the astronomical observations. Matching images to a stellar catalogue is a necessary prerequisite for determining the astrometric plate solution. \cite{groth1986pattern} had introduced the technique of looking for similar triangles in two catalogues which used the property of affine invariant under translation, rotation, magnification and inversion between two triangles.  \cite{valdes1995focas} used the ratio of sides as the invariant under the coordinate transformation to define its location in triangle space and accelerated the search process by pre-sorting the triangles. \cite{pal2006astrometry} described another method by culling the optimised triangles to accelerate the process.
\cite{murtagh1992new} proposed a method by characterizing a set of coordinates couples to decrease the computational complexity to $\mathcal{O}(N_1^2 N_2^2)$. \citep{mortari2000k} proposed a k-vector search method, a fast search algorithm based on monotone function and independent of data size, to reduce the matching time.\cite{tabur2007fast} proposed two fast matching algorithms based on triangle space and vector feature. The matching speed was accelerated by matching the sparse triangles with longest side and the shortest side, and the method based on vector feature reduced the computational complexity by constructing star pairs. \cite{RN540} applied the $k-d$ method to solve the more complicated  quadrilateral search problem and decreased the computational complexity dramatically from $\mathcal{O}(N_1^4 N_2^4)$ to $\mathcal{O}[(N_1^4 + N_2^4)logN_2]$. The computational complexity of our matching method based on statistical approach should be ${O}[(N_1^2 + N_2^2)log N_2]$ as the $k-d$ method and reduced the amount of calculation by culling the pre-sorting stars pairs. In practice, a priori knowledge of the telescope's focal length and the CCD detector's pixel size is common space, so the scale of the image is approximately known. Like the optimistic pattern matching described by \cite{tabur2007fast}, our high-speed matching process is accomplished under a slight loss of generality.

In section 2, we give an overview of the matching method based on the statistical result of star pairs. In section 3, we  analyze the performance of the method. We give our conclusions in section 4.
\section{Method}
\subsection{Construction of Star Pairs}
The first step of matching a image to a catalogue is to construct a list of sources ordered by gray values from the stellar detection and centring routine. The flat and bias corrections of the images are done before stellar detection and centring. We don't take all of the stars in the list to construct the star pairs.We only select the brightest $m$ stars from the image source list, denoted as $\mathscr{I}$. Similarly, we sort the stars from the catalogue by magnitude and extract the $n$ brightest stars, denoted as $\mathscr{R}$.  For m stars, $m(m-1)/2$ different star pairs could be obtained  by means of permutations and combinations. The equatorial coordinates of catalogue's stars should be converted to the tangent plane coordinates by the equation of \ref{Cc2Ic} \citep{green1985spherical}.  
\begin{equation}\label{Cc2Ic}
    \left\{
    \begin{aligned}
    \xi & = & \frac{\cos\delta\sin(\alpha-\alpha_0)}{\sin\delta\sin\delta_0+\cos\delta\cos\delta_0\cos(\alpha-\alpha_0)} 
    \\
    \eta & = & \frac{\sin\delta\cos\delta_0-\cos\delta\sin\delta_0\cos(\alpha-\alpha_0)}{\sin\delta\sin\delta_0+\cos\delta\cos\delta_0\cos(\alpha-\alpha_0)}
    \end{aligned}
    \right.
\end{equation}
where $\alpha$ and $\delta$ are the equatorial coordinates of catalogue's stars, $\alpha_0$ and $\delta_0$ are the coordinates of tangent plane's center point, $\xi$ and $\eta$ are the ideal coordinates in the tangent plane.  

\subsection{Searching for similar stars of the two lists}
The focal length of the telescope denoted as $F$, and the pixel size of CCD detector denoted as $\wp$. The scaling relations between the ideal coordinates of tangent plane and the measured coordinate of the image can be shown as:
$$ \rho = \frac{\wp}{F} $$
If the telescope adopt the binning mode, the pixel size should be changed accordingly.
The distance of star pair from the $\mathscr{I}$ list can be expressed as: 
$$
I_{ij} = \rho\times\sqrt{(x_i-x_j)^2+(y_i-y_j)^2}
$$
And the distance of star pair from the $\mathscr{R}$ list can be expressed as:
$$
R_{pq} = \sqrt{(x_p-x_q)^2+(y_p-y_q)^2}
$$
Then find the star pairs from the two lists that satisfy the following criteria:
\begin{equation}
 \mid I_{ij}-R_{pq}\mid < \delta F/F   
\end{equation}
where $\delta F$ is the uncertainty of the telescope's focal length. In practice we should increase the threshold appropriately. Because it's hard to obtain the accurate uncertainty of the focal length in real time, and the scaling relations $\rho$ contains centring errors, the tangent plane's center point error and so on. Each $I_{ij}$ might correspond to many $R_{ij}$.

\subsection{Find the rotation angle between the image and catalogue}
\label{rotation}
For each eligible star pair, we calculate the angle $\theta$, defined in the figure \ref{angle} :

\begin{figure}[!htbp]
  \centering
  \includegraphics[width=0.5\textwidth]{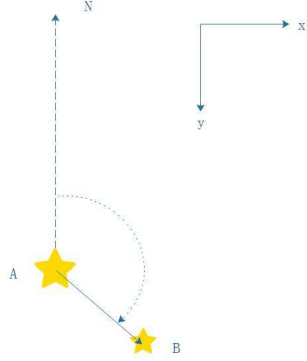}
  \caption{Angle of star pairs}
  \label{angle}
\end{figure}
 In figure \ref{angle}, the brighter star is marked as A and the other one is marked as B. The Angle is measured relative to north(although this is arbitrary) from $0^\circ$ to  $360^\circ$.

The rotation angle between the coordinates of tangent plane in celestial sphere and coordinates of image the is the difference of $\theta_I$  and $ \theta_R$
$$
\delta\theta = \theta_I -\theta_R
$$
If the $\delta\theta$ less than zero, we add $360^\circ$ to make the $\delta\theta$ in $0^\circ \sim 360^\circ$. We divide all the $\delta\theta$ into 360 bins($0^\circ \sim 360^\circ$), and count the amount of each bin. Looking for the most frequent of angle bins equals to find the mode of  statistical result of $\delta\theta$, this value is denoted as an initial value of rotation angle $\delta \theta_1$. This angle is a  estimated value. In order to obtain more precise angel, we need to find the subset of rotation angles $\delta \theta$ that are less then $1^\circ$ away from the initial angle $\delta\theta_1$.
    $$
    \mid \delta\theta - \delta\theta_1 \mid <1^\circ
    $$
The angels larger than 3 times standard deviations from the mean were excluded. Notice that the final mean angle of subset, denoted as $\delta \theta_2$, is more accurate than  $\delta \theta_1$. but $\delta \theta_2$ is not the most precise rotation angle of the affine transformation. 
\subsection{Determine the rough astrometric plate solution  }
\label{step4}
The affine transformation between the measured coordinates of image and ideal coordinates of tangent plane in celestial sphere could be roughly expressed  by a four-constant model as follows:
\begin{equation}
\label{4const}
    \left\{
  \begin{aligned}
  \xi & = \rho\cos\phi x - \rho\sin\phi x + c\\
  \eta & = \rho\sin\phi x +\rho\cos\phi y +d
  \end{aligned}
  \right.  
\end{equation}
where $\rho$ is the scale relations between two coordinate systems, and $\phi$ is the rotation angle.
In this case, the $\delta\theta_2$ is the adopted value of the rotation angle $\phi$. 
Applying the equation of \ref{4const}, we can obtain the translation of c and d. They are the average values after removing the outliers. Up to now, we get all the parameters of affine transformation.
\subsection{Marching for all of the stars and final verification  }
  Based on the initial plate solution obtained above, all stars detected from the image are transformed to equatorial coordinates using Equation \ref{Cc2Ic} and compare to the stars of reference catalogue to find their closest match. All possible star matching pairs are reduced by the least square under a four-constant or a six-constant model. A tolerance of 2.5$\sigma$ is used, the $\sigma$ is the residual of the least square. We then use the new plate solution to do the operation above again. The repeated operation are used to avoid some false matches caused by the imprecise initial plate solution in the last step of subsection \ref{step4}. We can use the residual of the least square to judge whether the image is successfully matched. Stars larger than 3 residual are considered to be mismatches.

\section{Experimental}
Several images are used to verify the reliability and speed of the method. The images are taken by the 1m telescope of Yunnan observatory. More details about the telescope and the CCD  detector are given in Table \ref{Specifications}. All of the images are taken by the $2*2$ binning acquisition mode. 

\begin{figure}[!htbp]
  \centering
  \includegraphics[width=\textwidth]{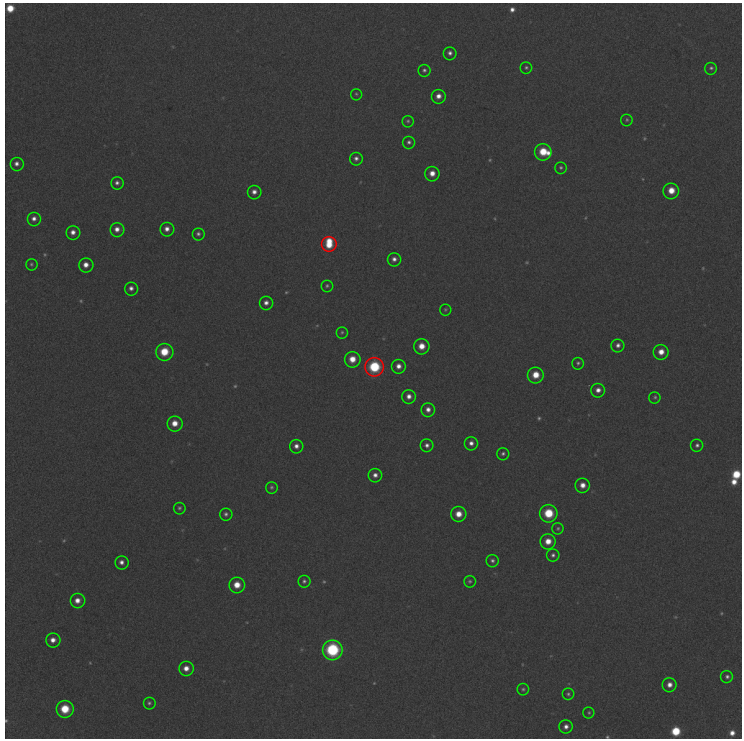}
  \caption{The successfully matched image}
  \label{matchingImage}
\end{figure}

\begin{table}
\begin{center}
\caption[]{ Specifications of 1-m Telescope and CCD Detector.}\label{Specifications}
 \begin{tabular}{lr}
  \hline\noalign{\smallskip}
  \hline\noalign{\smallskip}
Approximate focal length  & $1330 cm $ \\ 
F-ration  & $13$               \\
Diameter of primary mirror  & $100 cm $      \\
CCD field of view &   7.1 $^{'}\times$ 7.1$^{'}$ \\
Size of CCD array & $2048\times2048$ \\
Size of pixel & $13.5\mu\times13.5\mu$ \\
Approximate angular extent per pixel & $0.21^{''} $\\
  \noalign{\smallskip}\hline
\end{tabular}
\end{center}
\end{table}
The images are divided into two categories. One type of images contains asteroid and another contains M23 cluster. Reference stars are selected from the section of catalogue brighter than 18 magnitude of Gaia DR2\citep{lindegren2018gaia}. Figure \ref{matchingImage} is a successfully matched image. The green circles in the image is identified successfully, and the red ones are failed. The size of the circles is related to the magnitude of stars. The total number of stars in the catalogue is 217. Totally 77 stars are detected from the image, and 75 ones are matched to the catalogue stars successfully. Two of the failed stars in the image include an asteroid and a close binary. 

\subsection{ Robustness of the method}
\begin{figure}[!htbp]
    \centering
      \subfigure[]{
      \label{1a}
    \begin{minipage}[b]{0.5\textwidth}
      \includegraphics[width=\textwidth]{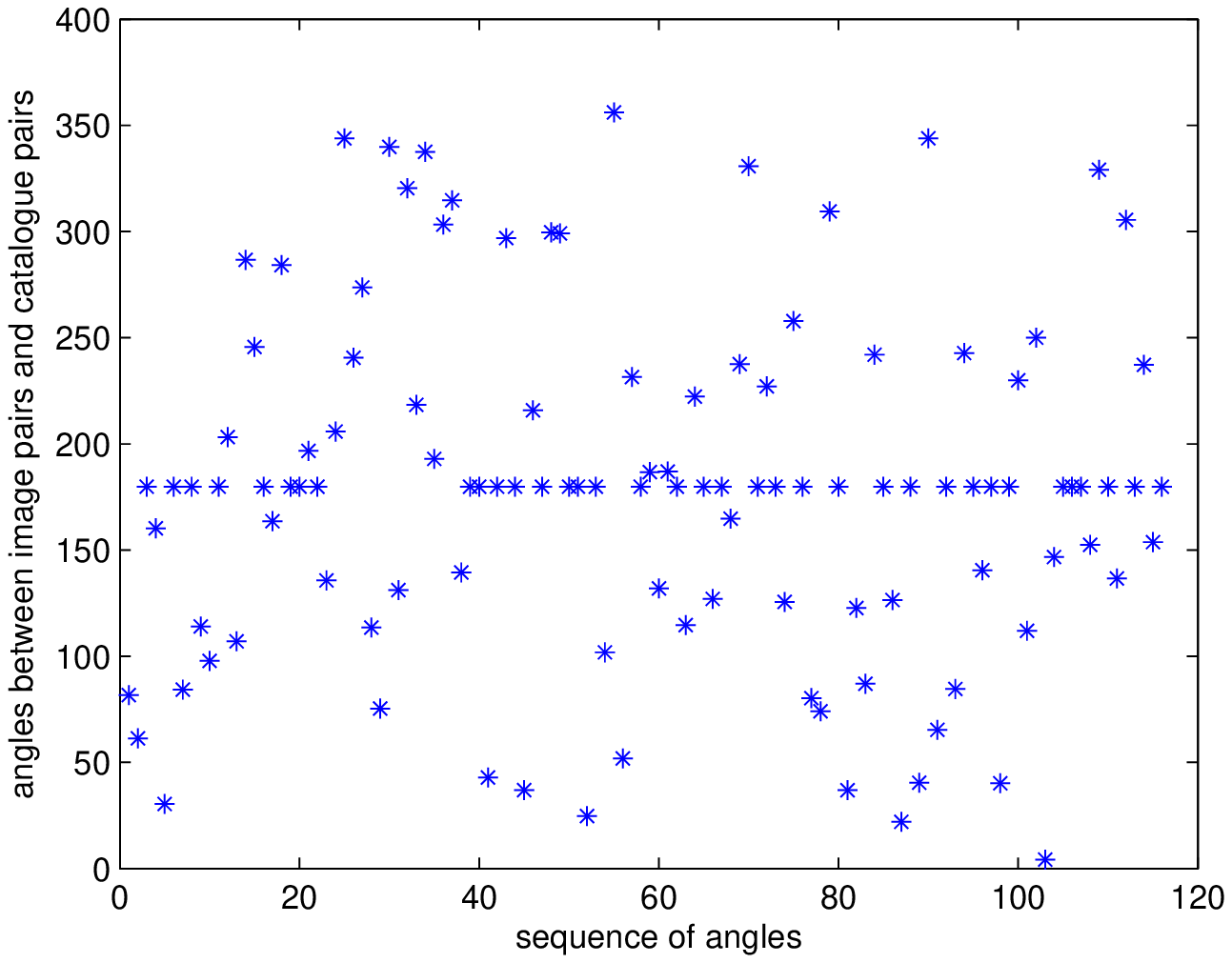}
    \end{minipage}}%
    ~
      \subfigure[]{
    \label{1b}
    \begin{minipage}[b]{0.5\textwidth}
      \includegraphics[width=\textwidth]{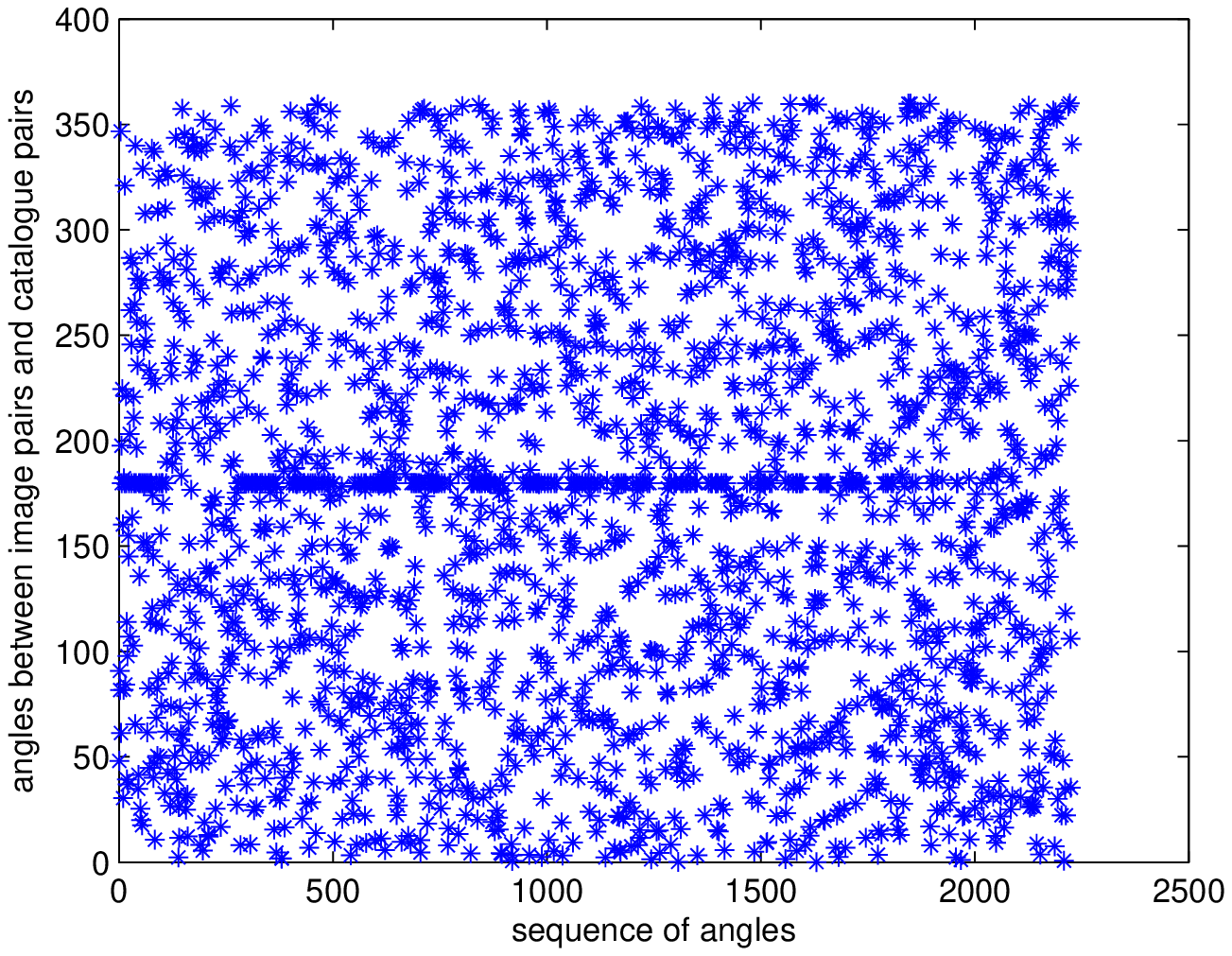}
    \end{minipage}}
      \subfigure[]{
            \label{1c}
    \begin{minipage}[b]{0.5\textwidth}
      \includegraphics[width=\textwidth]{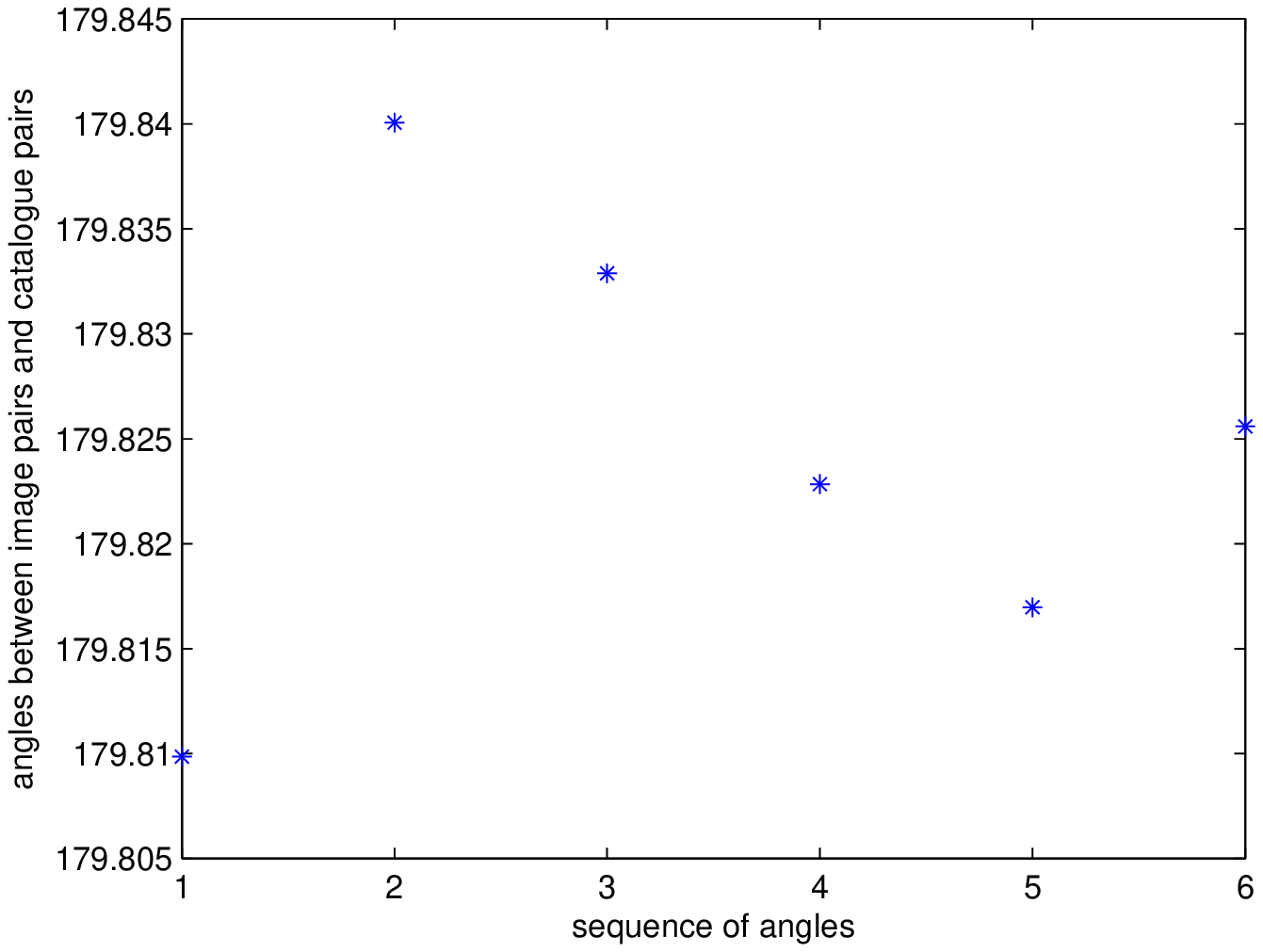}
    \end{minipage}}%
    ~
      \subfigure[]{
            \label{1d}
    \begin{minipage}[b]{0.5\textwidth}
      \includegraphics[width=\textwidth]{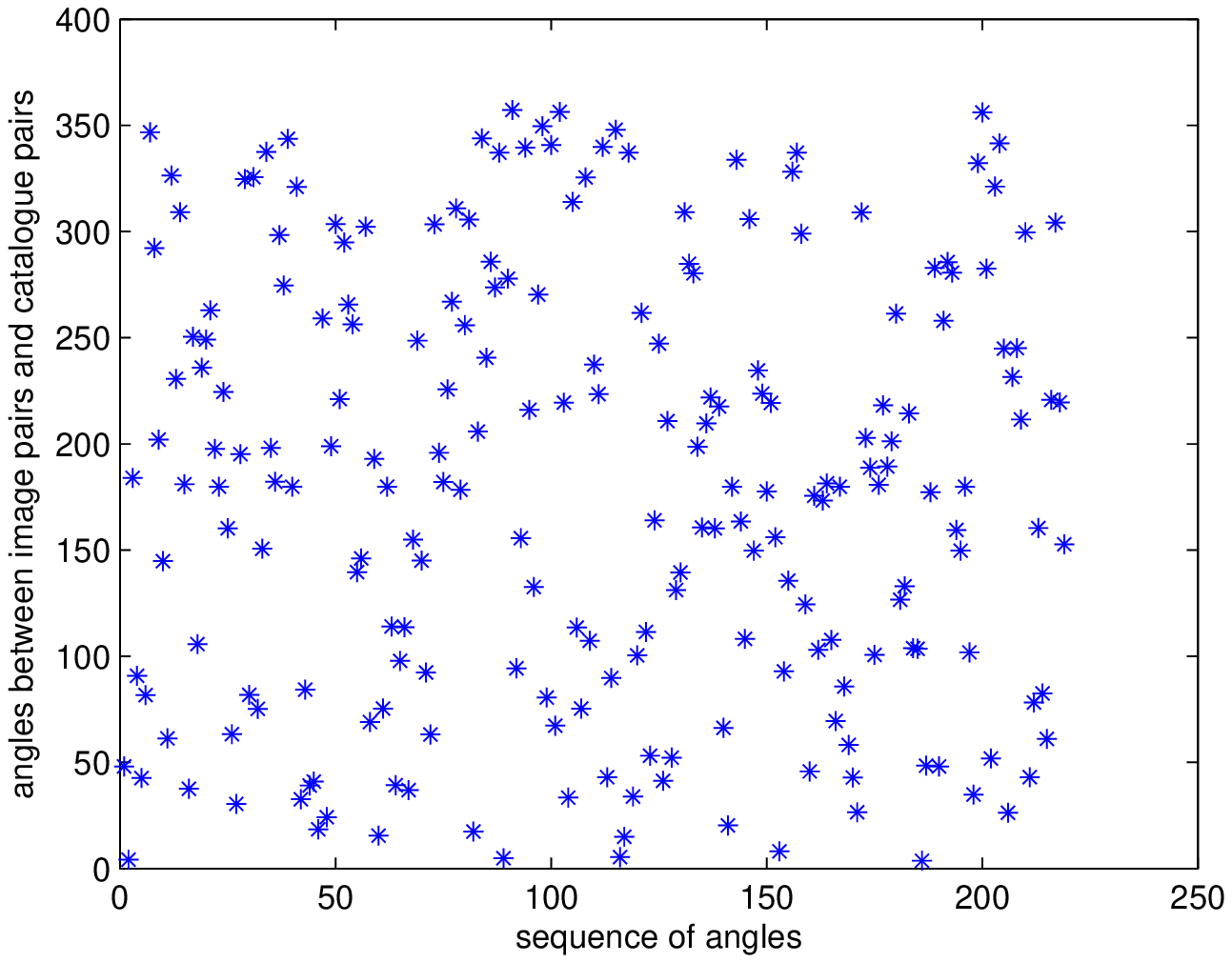}
    \end{minipage}}
    \caption{ distributions of $\delta\vartheta$ for the eligible stellar pairs. (a) m=10,n=30, (b) m=30,n=50, (c) m=5,n=10, (d) m=5,n=100.}
    \label{fig:oaspl}
\end{figure}

\begin{figure}[!htbp]
    \centering
    \subfigure[]{
          \label{2a}
    \begin{minipage}[b]{0.5\textwidth}
      \includegraphics[width=\textwidth]{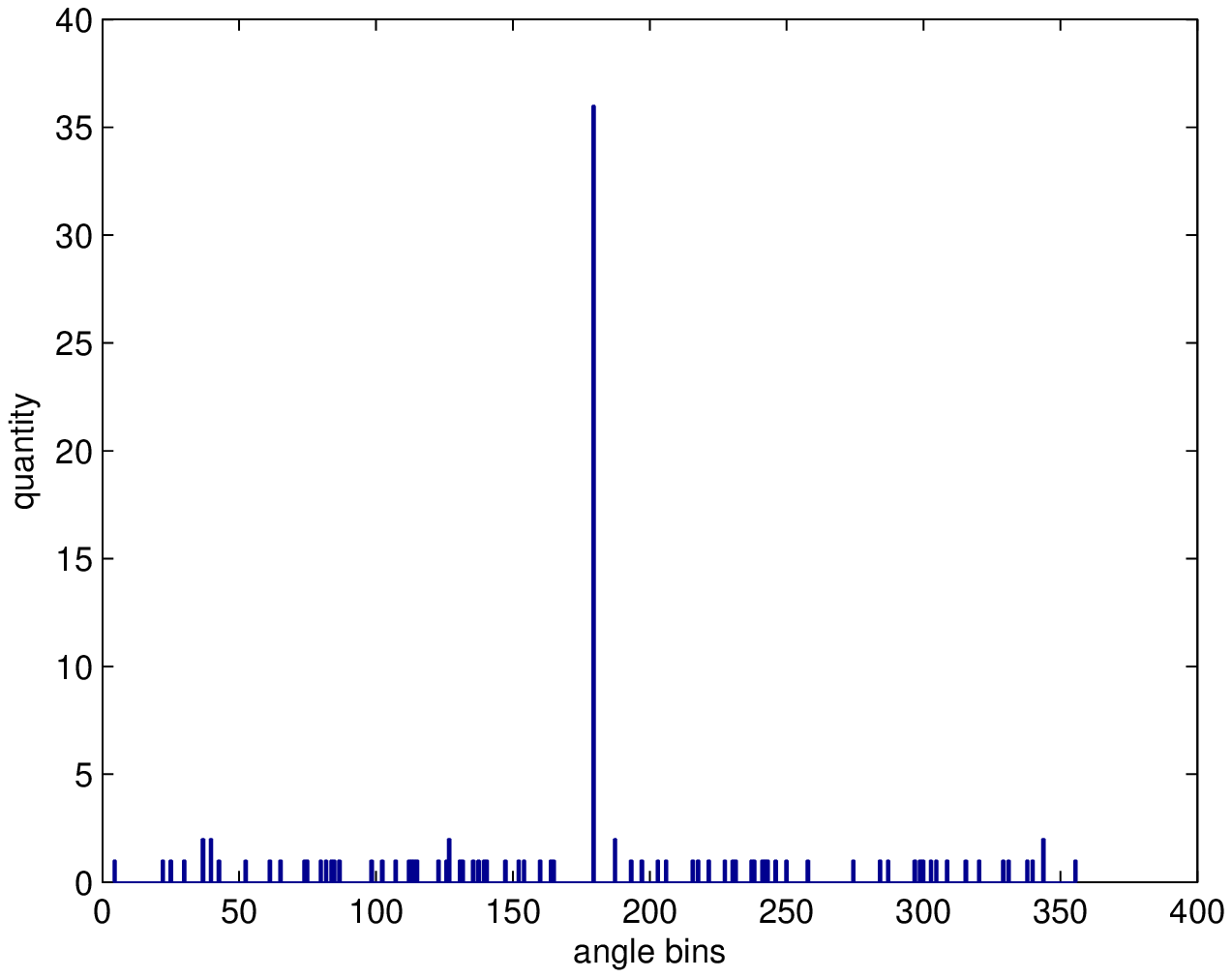}
    \end{minipage}}%
    ~
        \subfigure[]{
          \label{2b}
    \begin{minipage}[b]{0.5\textwidth}
      \includegraphics[width=\textwidth]{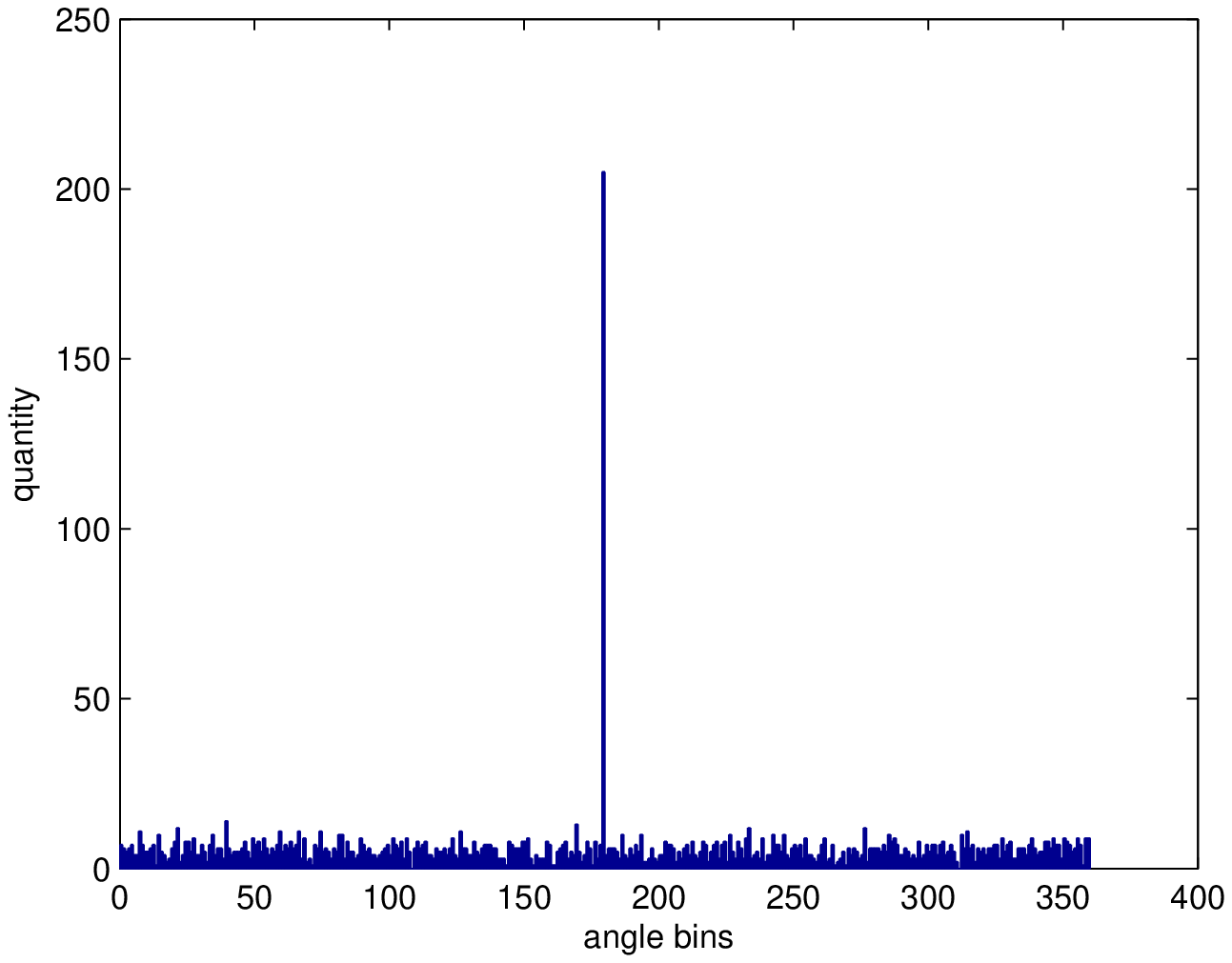}
    \end{minipage}}
        \subfigure[]{
                  \label{2c}
    \begin{minipage}[b]{0.5\textwidth}
      \includegraphics[width=\textwidth]{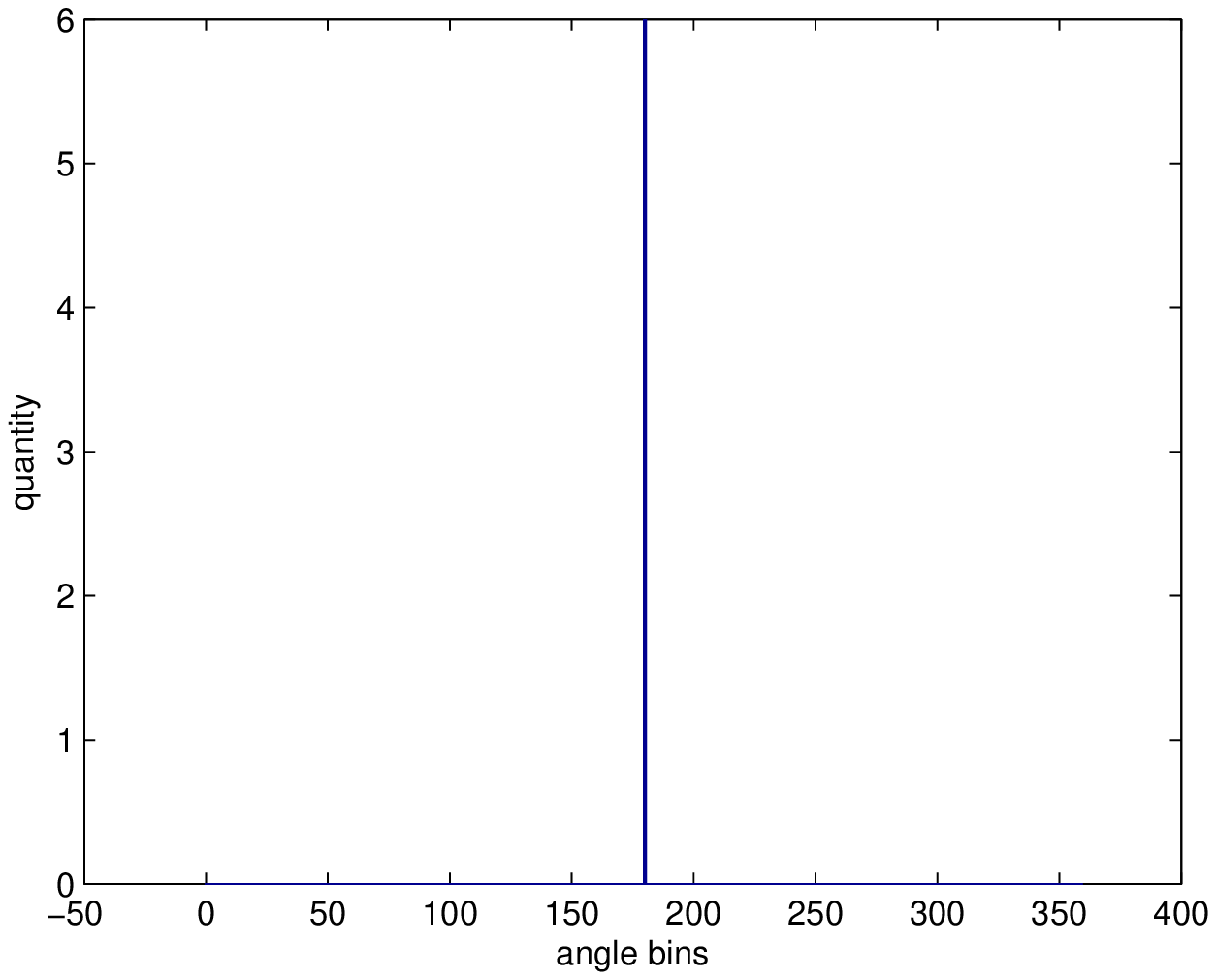}
    \end{minipage}}%
    ~
        \subfigure[]{
                  \label{2d}
    \begin{minipage}[b]{0.5\textwidth}
      \includegraphics[width=\textwidth]{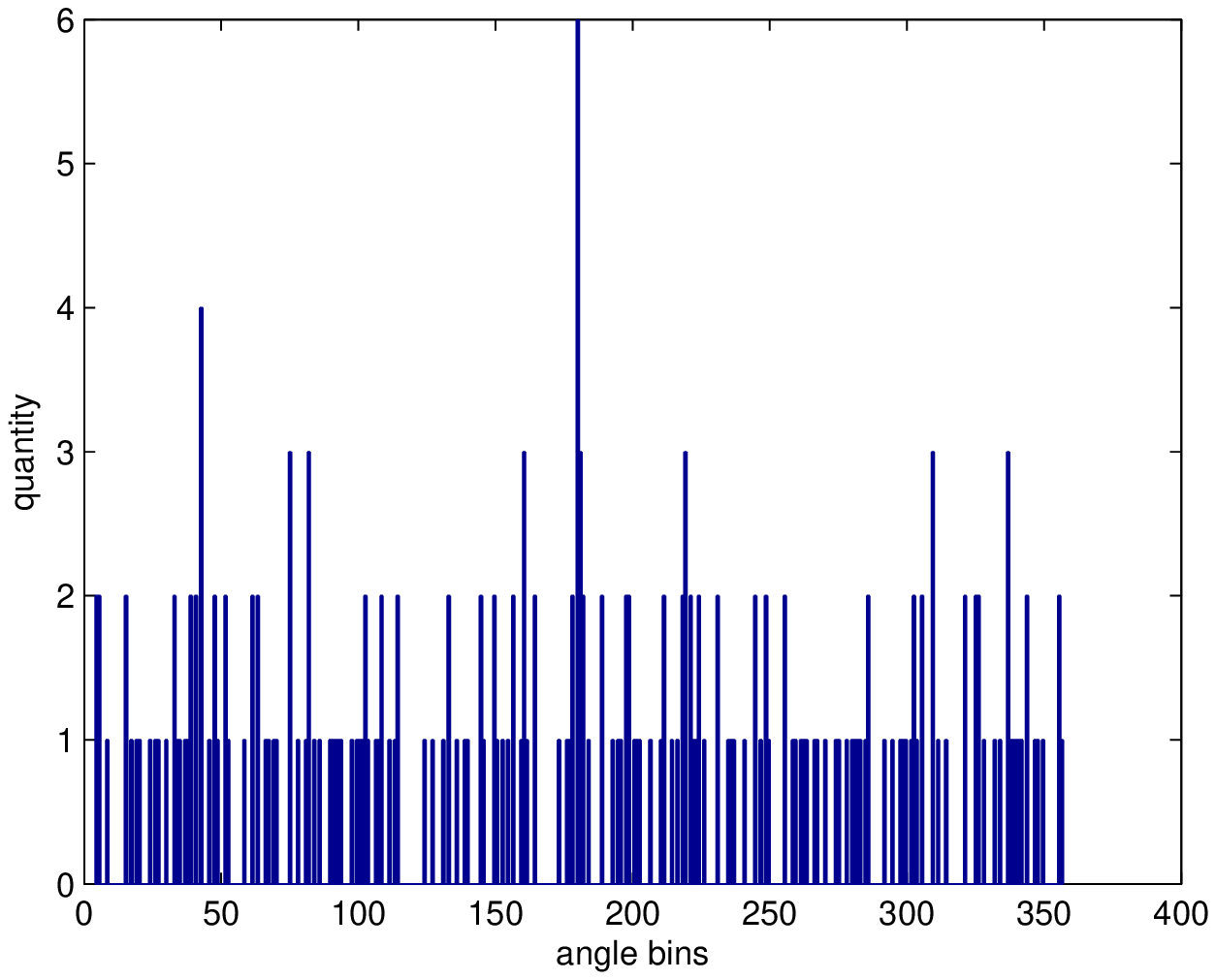}
    \end{minipage}}
    \caption{ Statistical histograms of $\delta\vartheta$ for the eligible stellar pairs. (a) m=10,n=30, (b) m=30,n=50, (c) m=5,n=10, (d) m=5,n=100.}
    \label{fig:oasp2}
\end{figure}

     Figure \ref{fig:oaspl} and Figure \ref{fig:oasp2} show the  statistical results of rotation angels based on different number of stars from the image and catalogue.  Obviously, the number of stars selected from the image m and the number stars selected from the catalogue n affects the speed of matching directly. Will it affects the success rate of matching? 

We select 10 stars from the the image and 30 from the catalogue to analyze the statistical result of rotation angle $\delta\theta$ derived from the subsection of \ref{rotation}. From the Figures of \ref{1a} and \ref{2a}, it can be seen that there exists a significant peak near $180^\circ$. The increase of both m and n like Figures \ref{1b} and \ref{2b}, or the decrease of m and n like Figure \ref{1c} and \ref{2c} would not affect the correct matching. There is still a significant spike around  $180^\circ$. In the extreme case, such as that m is 20 times n, it still match correctly even though the area of the catalog is only about twice as large as the CCD field of view.

\begin{figure}[!htbp]
  \centering
  \includegraphics[width=\textwidth]{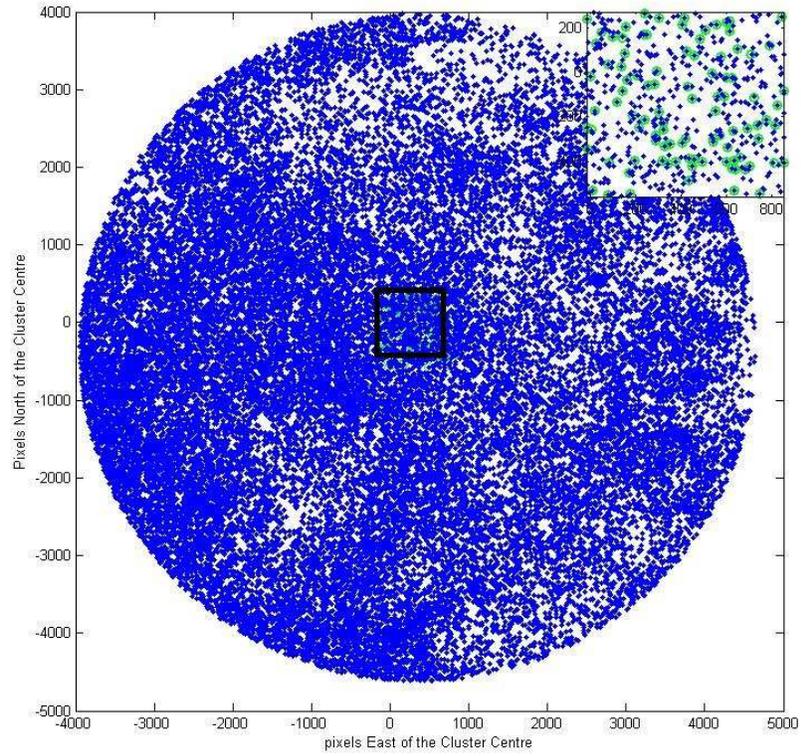}
  \caption{The positions of the  stars of the image in the cluster M23. The green circles are stars identified successfully and the blue points are points sources in the Gaia DR2 catalogue.The top right corner subgraph is a zoomed-in view of the black box area in the image. }
  \label{small}
\end{figure}

A difficult and exhaustive matching process is to search the right matching star in a dense star field. Figure \ref{small} shows the positions of the stars in the image of cluster M23. The image contains 135 stars, and the Gaia DR2 catalogue brighter than 18 magnitude around the center of M23 contains 33520 stars.

\subsection{Time consumption analysis}
\begin{figure}[!htbp]
    \centering
        \subfigure[]{
                  \label{3a}
    \begin{minipage}[b]{0.5\textwidth}
      \includegraphics[width=\textwidth]{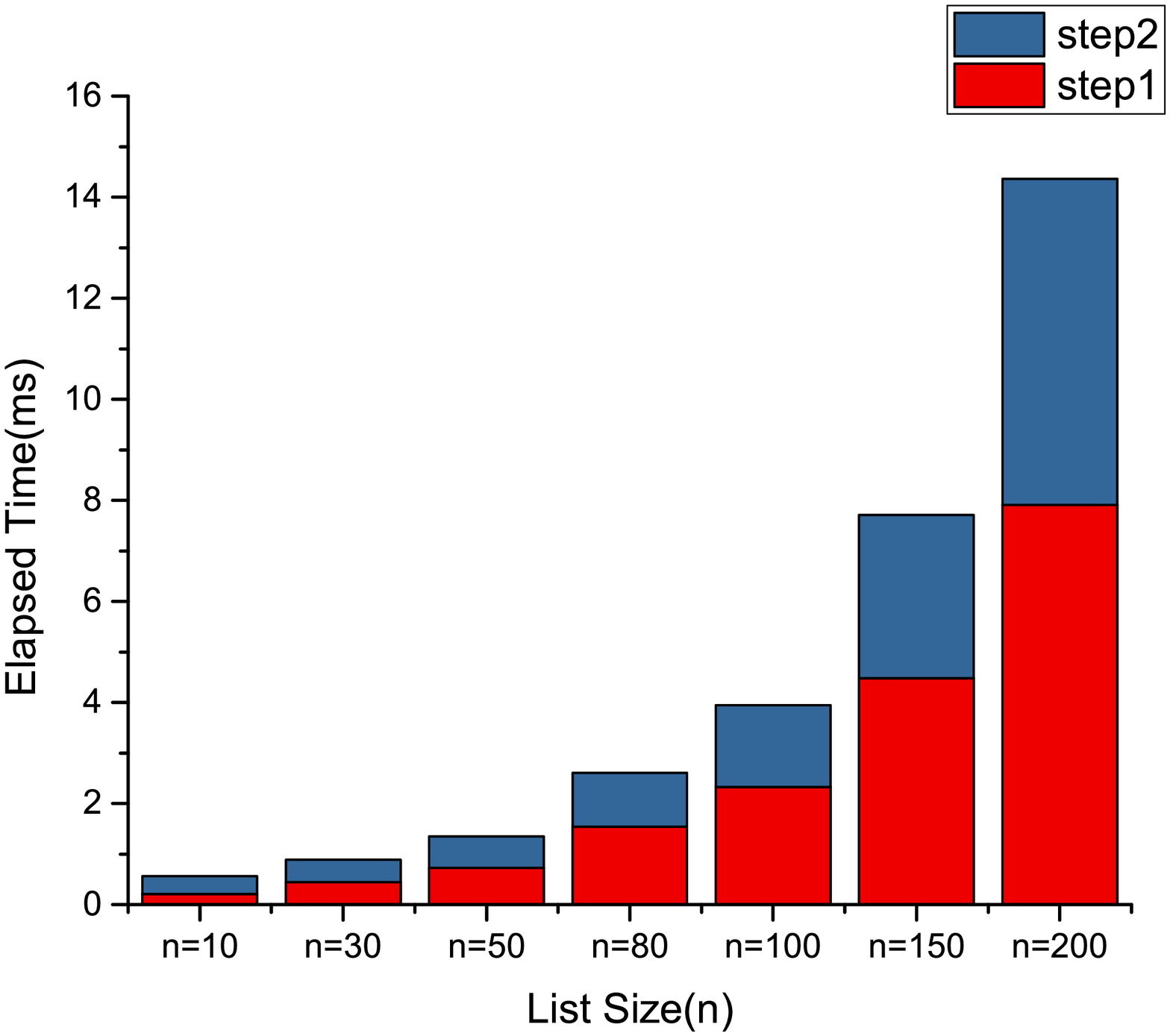}
    \end{minipage}}%
    ~
        \subfigure[]{
                  \label{3b}
    \begin{minipage}[b]{0.5\textwidth}
      \includegraphics[width=\textwidth]{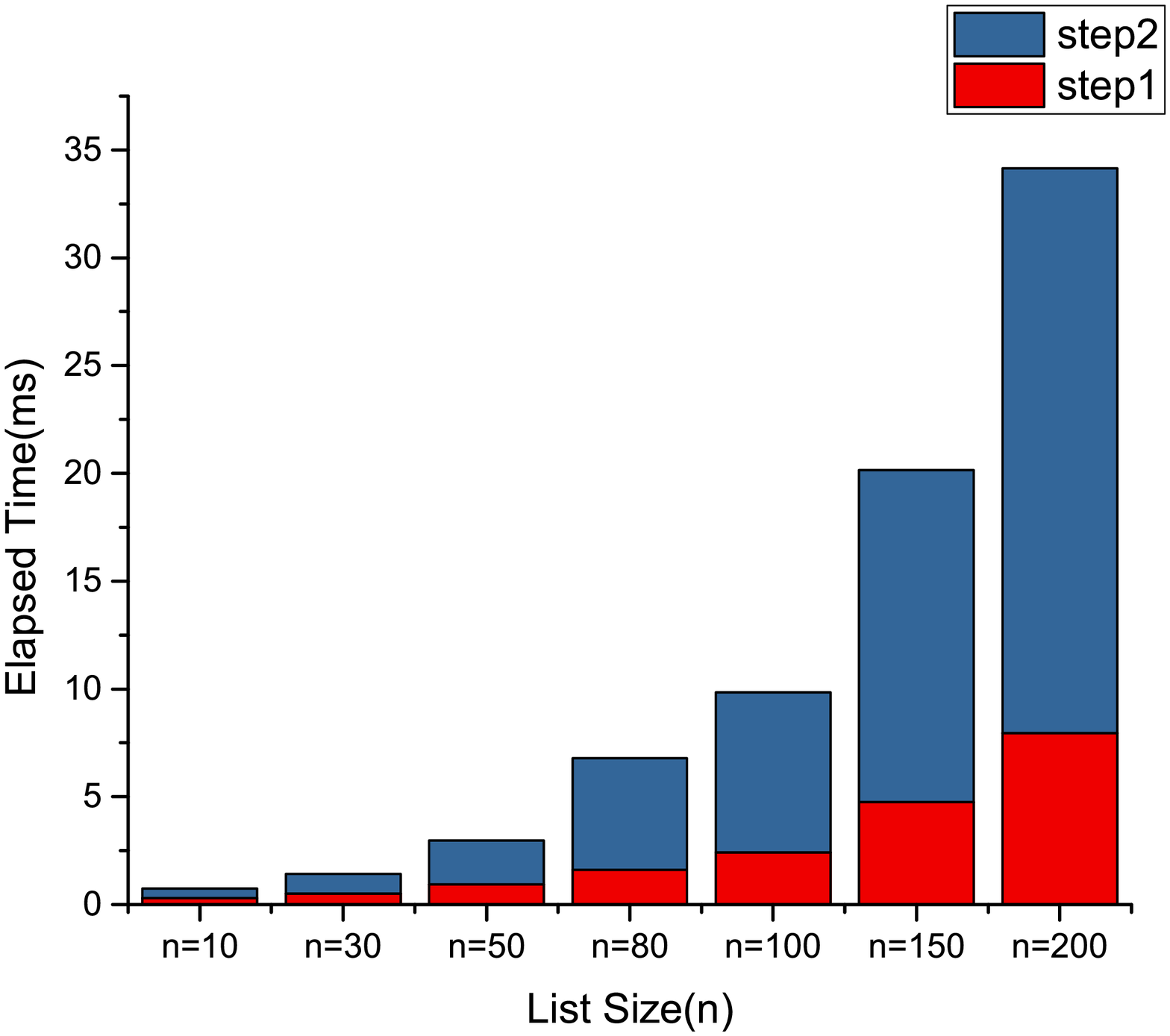}
    \end{minipage}}
    \caption{ Time Cost of Step1 and Step2. (a) The Case of List Size of $\mathscr{I}$ (m=5), (b) The Case of List Size of $\mathscr{I}$ (m=10).}
    \label{fig:m5m10}
\end{figure}

 The whole matching process can be divided into 3 steps.
 The first step is to construct the star pairs for calculating the angle distance and rotation angle of each star pair. The second step is to search the eligible star pairs from the two lists for determining the rough astrometric plate solution. The time required for two steps is independent to the total number of stars in the image and catalogue, and only relevant to the list sizes of $\mathscr{I}$ and $\mathscr{R}$. Separate timers are used to measure the performance of Step1 and Step2. Figure \ref{3a} and Figure \ref{3b} plot the time  costs of Step1 and Step2 in the cases of $m=5$ and $m=10$ respectively. The time costs are the average results over 1000 trials.
 
The combination of $m=5$ and $n=30$ can match the most images correctly. For the non-optimal condition of partly-overlapping caused by the pointing errors or some non-stellar targets,  the combination of $m=10$ and $n=50$ is used. and this combination can provide both high reliability and good performance\citep{tabur2007fast}. The first two steps can be completed in 3ms.

 The third step is to derive the exact plate solution and match all the stars from the image to catalogue. If there are M stars in the image and N star in the catalogue, the process of finding the nearest neighbour from the catalogue of step 3 can be sped up from ${O}(MN)$ to ${O}[(M+N)log M]$\citep{RN540,bentley1975multidimensional}.

\section{Conclusion}
This paper has outlined a efficient algorithm for matching images to a catalogue. For the most images, the star matching process can be completed in several millisecond. This method works at a slight loss of generality. We need to know the approximate focal length in advance, and attempt to match the field iteratively. It could be a viable solution when the focal length unknowns.
Fortunately, the focal length of telescope and the pixel size of CCD detector are known in most cases, and the change is very little. We can magnify the uncertainty of focal length to mend this matters. This method can deal with the geometric properties about translation, rotation, local distortion and extra/missing stars. But it should be noted that this method can not handle the case of image flipping, we must know the case and take countermeasures in advance.

\section*{Ackonwledgements}
We acknowledge the support of the staff of the 1-m telescope at Yunnan Observatory. This work is financially supported by the National Nature Science Foundation of China(grant nos.11503083,nos.11403101).This work has made use of data from the European Space Agency (ESA) mission
{\it Gaia} (\url{https://www.cosmos.esa.int/gaia}), processed by the {\it Gaia}
Data Processing and Analysis Consortium (DPAC,
\url{https://www.cosmos.esa.int/web/gaia/dpac/consortium}). Funding for the DPAC
has been provided by national institutions, in particular the institutions
participating in the {\it Gaia} Multilateral Agreement.

\section*{References}
\bibliography{mybibfile}

\end{document}